# Measuring Thermal Load in Fiber Amplifiers in the Presence of Transversal Mode Instabilities

Franz Beier,[1,2,*] Marco Plötner,[1] Bettina Sattler,[1] Fabian Stutzki,[1]
Till Walbaum,[1] Andreas Liem,[1] Nicoletta Haarlammert,[1]
Thomas Schreiber,[1] Ramona Eberhardt[1], and Andreas Tünnermann[1,2]

[1]Fraunhofer Institute for Applied Optics and Precision Engineering, Albert-Einstein-Str. 7, 07745 Jena, Germany
[2]Institute of Applied Physics, Abbe Center of Photonics, Friedrich-Schiller-University Jena, Albert-Einstein-Str. 15, 07745 Jena, Germany
*Corresponding author: franz.beier@iof.fraunhofer.de

**We report on detailed in-situ distributed temperature measurements inside a high power fiber amplifier. The deducted thermal load and the TMI-threshold of a commercial LMA fiber with 25 µm core and 400 µm cladding was measured at various seed wavelengths. By matching these results with detailed simulations we show that photodarkening has a negligible impact on the thermal load and, therefore, on the TMI threshold in this fiber.**

Power scaling of single-mode fiber amplifiers based on ytterbium-doped large mode area (LMA) fibers is limited by the onset of transversal mode instabilities (TMI) [1]: Above a certain average power, the beam quality is degraded by a sudden increase in higher-order mode content. The fundamental explanation describing this effect is mainly based on a thermally written longitudinal index grating that can become phase-matched to transfer energy between involved modes [2-5]. Over the years, a multitude of experiments has been carried out in order to deduce important influencing variables [6-9]. Nevertheless, due to changes in fiber properties and experimental parameters, it is difficult to compare these results. The influence of the thermal load, driven by quantum defect heating (QD) and photodarkening (PD), were discussed for nano-structured photonic crystal fibers [6, 10] and turn out as significant influence on the TMI-threshold.

Here, we present a detailed investigation of the wavelength dependence of the TMI-threshold for a commercially available Ytterbium-doped large-mode-area (LMA) step-index fiber with 25 µm core (NA=0.06) and 400 µm pump core that has been partly investigated before [11]. We will focus on the influence of the seed wavelength on the thermal load, determined from in-situ temperature measurements. In fact, the temperature measurement allows for an improvement of simulation parameters of a numerical rate equation simulation in steady state and, therefore, to state their relevance for the prediction of the TMI-threshold. As a result we will show that the influence of PD is negligible for the examined fiber type. Furthermore, the rate equation simulation with parameters fitted to the experimental thermal load is found to predict the TMI-threshold.

The 10 m long fiber amplifier under investigation was counter-pumped by a temperature stabilized diode laser at a wavelength of 976 nm as shown in Fig. 1. On the pump side, the fiber was fusion spliced to a fused silica end cap to improve the stability of the pump coupling. A pre-amplified tunable, phase modulated single frequency laser in the range of 1045 and 1075 nm with 10 W average power is used as seed source.

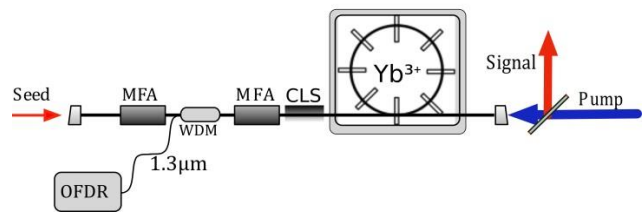

Fig. 1. Experimental fiber amplifier setup for temperature measurements. OFDR: Optical Frequency Domain Reflectometer, MFA: Mode Field Adapter, CLS: Clad Light Stripper. Fiber Length: 10m.

In order to investigate the temperature distribution inside the fiber core during high power operation, the Rayleigh-scattered part of a probe signal at 1.3 µm is monitored [12]. A fiber-based wavelength division multiplexer (WDM) combines and separates the light of the probe signal from an Optical Frequency Domain Reflectometer (OFDR) and the seed signal from each other [12, 13]. A clad light stripper was spliced between the active fiber and the second MFA to prevent the WDM from residual pump light. The spatial resolution of the temperature measurements is ~1 cm.

To ensure the reproducibility of the temperature measurements, care was taken on the thermal conductivity of the environment surrounding the fiber. Therefore, the fiber was inserted in a water bath containing non-moving water and was separated by polyethylene spacers at a bend radius of 6.5 cm. In order to validate the reproducibility and the physical feasibility of the longitudinal temperature measurements, several measurements have been performed. The quality of the temperature curves could be significantly improved compared to the results presented in [11]. In a first step, we ensured that the temperature measurement is not disturbed by any stress changes during operation. Furthermore, the temperature measurement was calibrated to the temperature of the water bath and the amplifier was operated over several hours to ensure that no significant time-dependent additional heat source such as photo darkening comes into play.

Fig. 2 shows typical curves of the temperature distribution at various pump powers and a signal wavelength of 1075 nm with regard to the steady state room temperature (294 K). The active fiber is localized between position 0 m (splice between Clad Light Stripper (CLS) and active fiber) and 10 m (pump end). In order to provide comparability of the curves and to calculate the average thermal load, an exponential fit was utilized. It should be noted that the temperature curves are disturbed beyond position 9 m by the forced cooling of the end cap and reflections of the probe signal. Therefore, this section was excluded from the exponential fit.

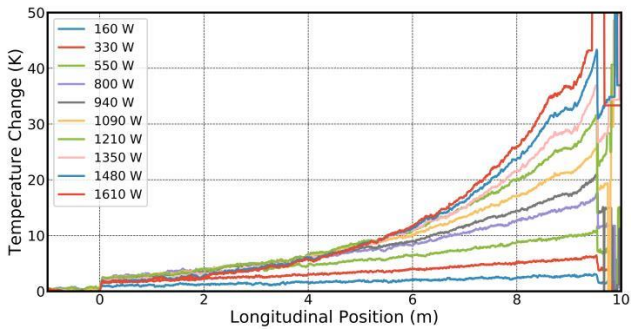

Fig. 2. Measured temperature change along the fiber amplifier for various output powers.

The temperature curves bend more strongly for higher power levels as the pump wavelength shifts towards 976 nm. Since the environment of the fiber is well known, the thermal load can be determined by solving the radial heat equation from the temperature curves. Assuming steady state operation, representing a constant longitudinal thermal load and a sufficient volume of the water bath, the heat equation can be solved as described in [14, 15]. The integration constants of the system can be determined by including the temperature boundary conditions resulting from the experimental setup [15]. A homogeneous heat flux was assumed as well as a homogeneous water temperature surrounding the fiber coating. Since the radially averaged temperature $\overline{T(z)}$ within the core at position z is the parameter determined from the OFDR-measurements, we integrated $T(z)$ according to eq. 7 in [15] over the radius within the core boundaries. Substituting the spatial solutions into each other, the volumetric heat flux $q$ results in

$$q(z) = \frac{\overline{T(z)}}{R_1^2} \cdot \left( \frac{1}{8a_1} + \frac{1}{a_2} ln\left(\frac{R_2}{R_1}\right) + \frac{1}{a_3} ln\left(\frac{R_3}{R_2}\right) \right)^{-1}, \quad (1)$$

where $\overline{T(z)}$ represents the radially averaged temperature of the core at a longitudinal position $z$, the thermal diffusivity $a_1, a_2$ and $a_3$ in the fiber core, cladding and coating with the radii $R_1, R_2$ and $R_3$, respectively. As the heat flux in longitudinal direction was neglected for the calculations, a local thermal load can be calculated by a scaling factor. The thermal properties of any cladding layers between the doped core and the coating were assumed to be constant. As an example, Fig. 3 shows the thermal load calculated from the measured temperatures with 1075 nm seed wavelength and 1100 W output power, resulting in 10.4 W/m average thermal load. At a fiber length of 10 m, this also fits very well with the total measured power losses (absorbed pump minus output power) of ~100 W.

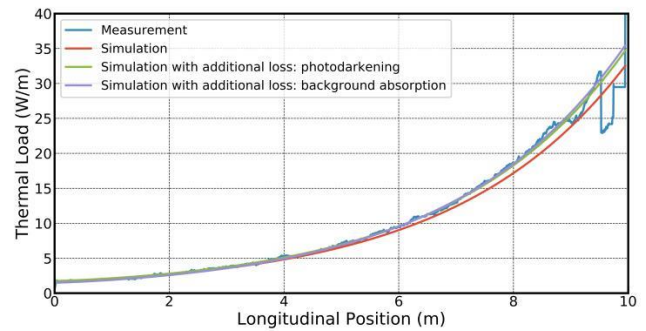

Fig. 3. Longitudinal distribution of the thermal load in comparison with a result from a rate equation simulation.

Implementing a rate equation simulation based on the experimental parameters, the theoretical thermal load can be calculated. Three significant sources of the thermal load can be identified: The quantum defect, photodarkening and background absorption of the fiber material. In a first step, only the quantum defect heating was considered in the simulation (Fig. 3, red curve). In a second step, the agreement of the simulation and the measurement has been improved by considering an additional loss mechanism leading to heat that is either a constant value (background absorption) or linearly dependent on the local population density [10]. The assumed photodarkening loss is depending on the excitation ratio along the fiber, but the simulation results (Fig. 3, green curve) is almost equivalent to a simulation done with a constant loss of around 8 dB/km for all studied wavelengths due to the almost constant excitation ratio in this backward pumped amplifier configuration (Fig. 3, purple curve).

We also ensured that any increased contribution of photodarkening with time can be excluded by repeating the experiments without any change in the temperature measurements within a total operation time of 20 hours. If photodarkening only has a minor contribution, most heat must be generated by quantum defect heating. This is demonstrated in the following experiments, that show that a change of the seed wavelength results in different temperature curves (exemplarily shown at an output power of 1350 W for a seed wavelength of 1045, 1055 and 1075 nm in Fig. 4). There, this additional loss of 8 dB/km is assumed in the comparing simulation and compared with QD-heating only.

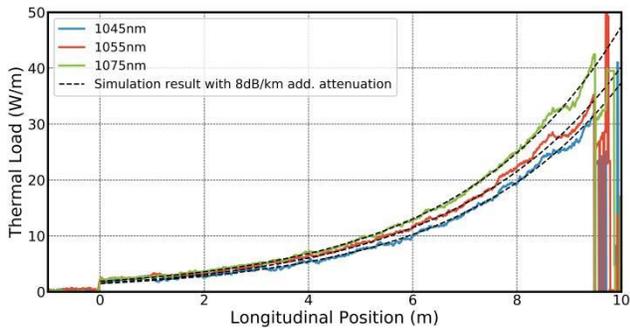

Fig. 4 Longitudinal thermal load for different seed wavelengths at 1350W output power in comparison to corresponding rate equation simulations (black).

Fig. 5 illustrates the retrieved average thermal load for different wavelengths at 1350 W output power. Obviously, the measured and simulated thermal load increases linearly with increasing seed wavelength as can be expected from quantum defect heating only. It should be noted that the pump wavelength stabilization of 976 nm was reached for every measured wavelength-dependent output power.

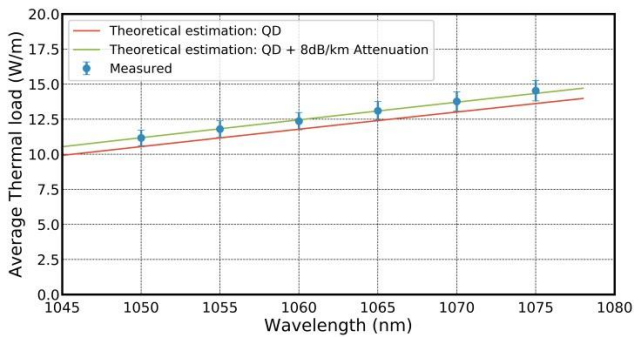

Fig. 5. Thermal load at a constant output power in dependence on the seed wavelength determined from the experiments in comparison with the simulation results for QD heating and QD heating with 8dB/km additional attenuation.

Extensive theoretical and experimental studies have shown that TMI are caused by an energy transfer between several longitudinal modes. A necessary condition for the transfer of energy is a phase shift between interference pattern of the modes and the thermally induced refractive index grating, which can be fulfilled by a moving temperature grating [2]. For this reason, longitudinal temperature measurements play an important role in the investigation of the effect and should enable us to draw conclusions about the thermal load and the physical processes that cause it. In the following, we will relate our measured heat loads to TMI threshold values. In our experiments, the TMI threshold is measured by imaging the near field output partly to a photo-diode and analyzing the normalized standard deviation σ against the output power [16]. A trace of 10s was split into several sections to create several values for the standard deviation σ since it varies over time. By plotting all values of σ over the corresponding output power, we get knowledge about the stability of the output signal. An exponential fit of the data and its derivative defined the TMI-threshold so far. For this fiber, we have observed that the steepness changes for different wavelengths as shown in Fig. 6. Therefore, we define the TMI threshold as power level with fivefold increase of the calculated average standard deviation much below the threshold similar to [17]. In Fig. 6, the TMI thresholds were found at an output power of 1250W and 1350W for the red and the blue values, respectively. In this manner, the TMI threshold of the fiber under investigation was measured for different seed wavelengths.

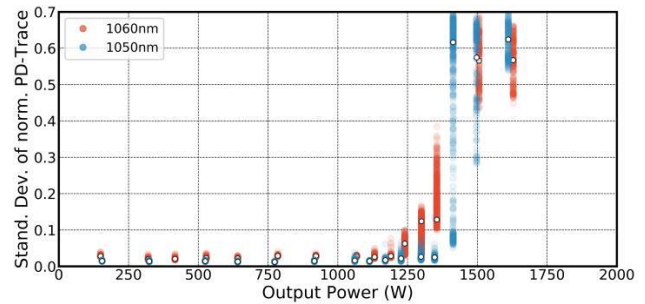

Fig. 6. Standard deviation of the normalized photo-diode time traces for two different wavelengths. The white markers represent the average standard deviation.

Fig. 7 shows the determined TMI thresholds between 1050 and 1075 nm and in addition for different bend diameters. Before we ensured that the temperature distribution is unchanged when bending the fiber. Overall, the TMI threshold is increased by the use of shorter seed wavelengths from around 1000 W at 1075 nm up to 1370 W at 1050 nm. Since the thermal load has been discussed as one driving parameter for TMI in the literature, we determined the average thermal load for 1075 nm and shifted the seed wavelength in the simulation afterwards, while the pump power was increased until an identical thermal load of 11.2 W/m in average and 37.2 W/m as maximum value are obtained [2, 6].

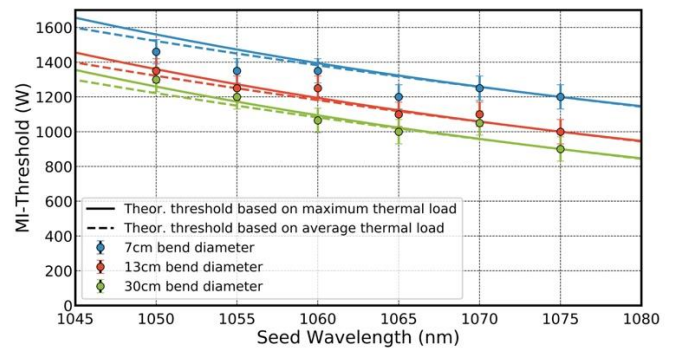

Fig. 7. Transversal mode instability threshold in dependence on the seed wavelength in comparison to the theoretical curve of constant average and maximum thermal load.

Fig. 7 shows the curves calculated in this way for the average and the maximum thermal load. However, this thermal load is not a fundamental constant for this or other fibers, but in our case is valid for the given bending diameter and system layout. Furthermore, when the bend diameter is decreased it can be seen that the TMI threshold increases. Two significant findings can be obtained from Fig. 3, Fig. 4, Fig. 5 and Fig. 7. Firstly, the thermal load of such an LMA-fiber amplifier can be described very exactly by rate equation simulations including the absorption parameters obtained from in situ temperature measurements. On the one hand, this is important as these very low photodarkening values are hard to measure especially with this low-NA fiber. On the other hand, background loss measurements contain contributions that are not related to heat causes, for instance, scattering losses. Secondly, the measured TMI-thresholds within the scope of an LMA-fiber are in acceptable accordance with a maintained thermal load, whether considering the average or the maximum value. Unfortunately, the longitudinal resolution of this temperature measurement technique (~1cm) is insufficient (despite the fact that the Rayleigh resolution is only 10 μm) to directly see the expected index grating predicted by theory [18, 2]. Besides this, the moving temperature grating might be averaged in time since every Rayleigh-scattering measurement takes several seconds.

In conclusion, we presented a detailed investigation of TMI thresholds with respect to the seed wavelength for a step-index LMA-fiber with 25μm core and 400μm cladding diameter at a fiber length of 10 m and at various bend diameters. In addition to the TMI-threshold measurements, all results are complemented by in-situ core temperature measurements during high power operation. Finally, we conclude that in the fiber under investigation, the additional loss contributing to heat is very low with a value of only 8 dB / km and is therefore negligible in contrast to the regime found in other fibers. Thus the TMI threshold of the fiber seems to be driven mainly by quantum defect heating with all other parameters fixed and an influence of photo-darkening and its dependence on inversion and seed-wavelength could not be observed. To the best of our knowledge this is the first systematic in-situ measurement of core temperatures in a high power amplifier fiber and this technique is ideally suited for detailed fiber amplifier analysis.

**Funding.** BMBF (13N13652), EMPIR (14IND13), EU project MIMAS (670557), The Fraunhofer and Max Planck cooperation program (PowerQuant), State of Thuringia supported by EU programs EFRE and ESF (2015FOR0017, 13030-715, 2015FGR0107, B715-11011)

## References

1. T. Eidam, C. Wirth, C. Jauregui, F. Stutzki, F. Jansen, H. Otto, O. Schmidt, T. Schreiber, J. Limpert, and A. Tünnermann, "Experimental observations of the threshold-like onset of mode instabilities in high power fiber amplifiers," Opt. Express **19**, 13218-13224 (2011).
2. C. Jauregui, T. Eidam, H. Otto, F. Stutzki, F. Jansen, J. Limpert, and A. Tünnermann, "Physical origin of mode instabilities in high-power fiber laser systems," Opt. Express **20**, 12912-12925 (2012).
3. A. Smith and J. Smith, "Steady-periodic method for modeling mode instability in fiber amplifiers," Opt. Express **21**, 2606-2623 (2013).
4. B. Ward, C. Robin, and I. Dajani, "Origin of thermal modal instabilities in large mode area fiber amplifiers," Opt. Express **20**, 11407-11422 (2012)
5. S. Naderi, I. Dajani, T. Madden, and C. Robin, "Investigations of modal instabilities in fiber amplifiers through detailed numerical simulations," Opt. Express **21**, 16111-16129 (2013).
6. H. Otto, N. Modsching, C. Jauregui, J. Limpert, and A. Tünnermann, "Impact of photodarkening on the mode instability threshold," Opt. Express **23**, 15265-15277 (2015).
7. N. Haarlammert, O. de Vries, A. Liem, A. Kliner, T. Peschel, T. Schreiber, R. Eberhardt, and A. Tünnermann, "Build up and decay of mode instability in a high power fiber amplifier," Opt. Express **20**, 13274-13283 (2012).
8. R. Tao, P. Ma, X. Wang, P. Zhou, and Z. Liu, "1.3kW monolithic linearly polarized single-mode master oscillator power amplifier and strategies for mitigating mode instabilities," Photon. Res. **3**, 86-93 (2015).
9. F. Beier, C. Hupel, J. Nold, S. Kuhn, S. Hein, J. Ihring, B. Sattler, N. Haarlammert, T. Schreiber, R. Eberhardt, and A. Tünnermann, "Narrow linewidth, single mode 3 kW average power from a directly diode pumped ytterbium-doped low NA fiber amplifier," Opt. Express **24**, 6011-6020 (2016)
10. C. Jauregui, H. Otto, F. Stutzki, J. Limpert, and A. Tünnermann, "Simplified modelling the mode instability threshold of high power fiber amplifiers in the presence of photodarkening," Opt. Express **23**, 20203-20218 (2015).
11. F. Beier ; M. Heinzig ; Bettina Sattler ; Till Walbaum ; N. Haarlammert ; T. Schreiber ; R. Eberhardt and A. Tünnermann, " Temperature measurements in an ytterbium fiber amplifier up to the mode instability threshold ", Proc. SPIE **9728**, Fiber Lasers XIII: Technology, Systems, and Applications, 97282P (March 11, 2016)
12. B. Soller, M. Wolfe, and M. Froggatt, "Polarization Resolved Measurement of Rayleigh Backscatter in Fiber-Optic Components," in Optical Fiber Communication Conference and Exposition and The National Fiber Optic Engineers Conference, Technical Digest (CD) (Optical Society of America, 2005), paper NWD3.
13. B. Soller, D. Gifford, M. Wolfe, and M. Froggatt, "High resolution optical frequency domain reflectometry for characterization of components and assemblies," Opt. Express **13**, 666-674 (2005).
14. J. Crank, "The mathematics of diffusion", J. Crank Clarendon Press Oxford, 1975
15. D. C. Brown and H. J. Hoffman, "Thermal Stress and Thermo-Optic Effects in High Average Power Double-Clad Silica Fiber Lasers," IEEE J. Quantum Electron. **37**, 2, (2001).
16. H. Otto, F. Stutzki, F. Jansen, T. Eidam, C. Jauregui, J. Limpert, and A. Tünnermann, "Temporal dynamics of mode instabilities in high-power fiber lasers and amplifiers," Opt. Express **20**, 15710-15722 (2012).
17. M. Johansen, M. Laurila, M. Maack, D. Noordegraaf, C. Jakobsen, T. Alkeskjold, and J. Lægsgaard, "Frequency resolved transverse mode instability in rod fiber amplifiers," Opt. Express **21**, 21847-21856 (2013).
18. A. Smith and J. Smith, "Mode instability in high power fiber amplifiers," Opt. Express 19, 10180-10192 (2011).